\documentclass[iop]{emulateapj}

\usepackage{times}
\usepackage{amssymb}
\usepackage{rotating}
\usepackage{epsfig}

\shorttitle{Minimum X-ray source size in AGN}
\shortauthors{Dov\v{c}iak \& Done}

\def\mnras{MNRAS}
\def\apj{ApJ
}\def\apjl{ApJL}
\def\apjs{ApJS}
\def\aap{A\&{A}}

\def\pasj{PASJ}
\def\nat{Nature}
\def\msun{M$_\odot$}
\def\dd{{\rm d}}

\begin{document}

\title{Minimum X-ray source size for a lamp-post corona in light-bending
  models for AGN} 
  
\author{M.~Dov\v{c}iak,\altaffilmark{1,3}}
\author{C.~Done\altaffilmark{2,4}}

\altaffiltext{1}{Astronomical Institute, Academy of Sciences of the Czech 
Republic, Bo\v{c}n\'{\i}~II 1401, CZ-14100 Prague, Czech~Republic}
\altaffiltext{2}{Centre for Extragalactic Astronomy, Department of Physics, 
University of Durham, South Road, Durham DH1 3LE, United Kingdom}
\altaffiltext{3}{Michal.Dovciak@asu.cas.cz}
\altaffiltext{4}{Chris.Done@durham.ac.uk}
  
\label{firstpage}

\begin{abstract}

The 'lamppost' model is often used to describe the X-ray source
geometry in AGN, where an infinitesimal point source is located on
the black hole spin axis.  This is especially invoked for Narrow Line
Seyfert 1 (NLS1) galaxies, where an extremely broad iron line seen in
episodes of low X-ray flux can both be explained by extremely strong
relativistic effects as the source approaches the black hole
horizon.The most extreme spectrum seen from the NLS1 1H0707-495
requires that the source is less than 1Rg above the event horizon 
in this geometry. 
However, the source must also be large enough to intercept
sufficient seed photons from the disk to make the hard X-ray Compton
continuum which produces the observed iron line/reflected spectrum.
We use a fully relativistic ray tracing code to show that this implies that  the
source must be substantially larger than 1Rg in 1H0707-495 if the disk is the 
source of seed photons. 
Hence the source cannot fit as close as 1Rg to the horizon, so the
observed spectrum and variability are not formed purely by effects of 
strong gravity but probably also by changes in corona and inner accretion 
flow geometry.
\end{abstract}

\keywords{accretion, accretion disks --- black hole physics --- galaxies: active
--- relativistic processes 
--- X-rays: galaxies}

\section{Introduction} \label{sec:introduction}

Very broad iron K$\alpha$ lines are seen in the X-ray spectra of some
Active Galactic Nuclei (AGN), indicating that the X-ray illuminated
disk extends deep into the gravitational potential well of the black
hole.  The broadest lines, indicating the most extreme velocities and
strongest gravitational fields, are seen in Narrow Line Seyfert 1
(NLS1) galaxies such as 1H0707-495 and IRAS13224-3809. Spectral
fitting with a power law and its reflection over the 2-10 keV range
requires that the black hole has high spin, so that the innermost
stable orbit of the material in the disk is below 2Rg, and that
the line emissivity is strongly centrally peaked onto this inner disk
edge, and that the spectrum is dominated by the reflected emission
rather than the intrinsic continuum. These properties were first seen
in MCG-6-30-15 \citep{wilm01} but are much more extreme in
1H0707-495 \citep{fabi09,zogh10} and IRAS13224-3809
\citep{pont10,chia15}. This behavior can be produced
from the light-bending model \citep{mini04}, where a small
'lamppost' source moves in the vicinity of the black hole changing its distance 
from the center.
Strong relativistic effects can simultaneously boost the continuum as
seen by the very inner edge of the disk, and reduce its outwards flux
towards the observer when the source is very close to the event
horizon of a high spin black hole. This results in a more
or less constant reflection intensity, but with increasing smearing
from the increasingly centrally concentrated illumination, 
while the observed primary power law emission drops \citep{mini04}.

This lamp-post geometry gives a successful framework in which to
interpret the spectra and spectral variability seen in these extreme
NLS1 \citep{fabi09, pont10, zogh10,chia15}. 
However, most modeling to date has assumed a toy model
of an infinitesimal (point) source. This is the simplest case for
general relativistic ray tracing calculations, and \citealt{daus13}
shows that a more physical vertically extended region can be well
approximated by a point source at some effective intermediate height.

Here we try to set a lower limit on the size of an X-ray source on the
spin axis from assuming that its emission is from Compton scattering
of seed photons from the disk. Compton scattering conserves photon
number so the corona must intercept at least as many seed photons as
are required to form the observed reflected emission. We include both
direct disk emission and the flux from re-processing of the
illuminating source, and self consistently calculate the resulting
photon density on the black hole spin axis including full general
relativistic effects. We estimate this minimum radius for a spherical
source, and find that it is compact but not tiny for an intrinsic
$L_{\rm X}/L_{\rm disk}\sim 0.03$, as is typically observed in NLS1. This source
can only fit within the space if its height above the horizon is
larger than its radius (Boyer Lindquist coordinates) which limits the
height to $>10R_{\rm g}$ for an on-axis source for the steepest observed
spectra ($\Gamma=3$). 

However, tailoring the model to the mean spectrum of the most extreme
NLS1, 1H0707-495, reveals a much stronger inconsistency. The centrally
concentrated emissivity \citep[radial index of $\sim 6.6$:][]{fabi12}
and observed high reflected fraction \citep[$\Omega/2\pi>10$:][]{zogh10}
are consistent with a source at $h< 2$. However, at
this low height the observed $L_{\rm X,obs}$ is dramatically reduced from the
intrinsic one, so the intrinsic source $L_{\rm X}\sim1.9L_{\rm Edd}$. This 
requires a source size which is much larger than for 
$L_{\rm X}\sim0.03L_{\rm Edd}$, so this cannot fit within $h<2R_{\rm g}$
for any spectral index. Some other geometry is instead required. We note that a
corona which extends radially over this disk \citep[e.g.][]{wilk13} will 
see many more photons than an on axis source. 

Either the extent of the relativistic effects are overestimated from the 
spectra (e.g. the steepness of the radial emissivity profiles could be biased to 
higher values by the assumption of a constant ionization reflected spectrum 
\citep{svob12} or by the effects of complex absorption \citep[e.g.][]{tana04, 
mill13}), or relativistic effects alone do not 
control both the spectrum and variability of the source e.g the deep dips cannot 
be caused by light-bending in the radially extended corona model of 
\cite{wilk14}.

\section{Seed photons and Comptonization: non-relativistic approach}

NLS1 are typically lower mass, higher mass accretion rate AGN than
standard broad line Seyfert 1s. We show all calculations for mass of
$10^{7}$\msun, $L_{\rm disk}\sim L_{\rm Edd}$, with maximal spin, $a_*=0.998$.
We assume an inclination angle of $30^\circ$ to the spin axis but this
makes little difference to the results. All size scales $r,h,\dd h$
(radii on the disk, height of the corona, radial size of the corona)
are given in units of the gravitational radius, so $R=rR_{\rm g}$, 
$H=hR_{\rm g}$, and $\dd H=\dd h\,R_{\rm g}$.

We first show results in a 'Newtonian' framework, where we assume that
there is no change in photon energy or rate between the disk and the
lamp-post, or between these and the observer at infinity.  We also
assume that light travels in straight lines, but we use the full
Novikov-Thorne emissivity for ease of comparison with the fully
relativistic approach in the next section. This gives the surface
temperature $T(r)$ in the comoving disk frame, which peaks at
$T_{\rm max}$ at $r=r_{\rm max}\sim 1.6$ for $a_*=0.998$.  This inner
disk temperature is high enough to produce substantial emission in the
observable soft X-ray regime, especially if electron scattering within
the disk acts as expected, giving a color temperature correction
$f_{\rm col}=2.2-2.6$ \citep{ross92,done12,done13}. 
We use $f_{\rm col}=2.4$ in all results shown.

The key determinant of the source size is the ratio of photons
required for Comptonization compared to the photons available, which
in turn is set by $L_{\rm X}/L_{\rm disk}$. Observations show that this is
generally small in NLS1s, at around $0.02$ \citep{vasu07,jin12,done12}.
For ease of comparison to the full
relativistic calculation in the next section we here include the
effect of redshift from the source to the observer. For a source at
height $h=10$ then an observed $L_{\rm X,obs}=0.02L_{\rm Edd}$ corresponds to a
luminosity in the lamp-post frame of $L_{\rm X}=0.031L_{\rm Edd}$. Following the
light bending model, we assume that this is constant as a function of
source height.

The coronal emission is soft, with power law photon index $\Gamma=2-3$
\citep{shem06,vasu07,jin12,done12}
between a low energy cutoff set by the seed photon
temperature and a high energy cutoff set by the electron temperature
which we fix at 100~keV in the lamppost frame. For a steep spectrum,
the number of photons is critically dependent on the low energy
cutoff, which is at $2-4kT_{\rm BB}$, where $T_{\rm BB}$ is the temperature of 
the seed photons in the lamp-post frame assuming they have a blackbody
distribution \citep[e.g.][]{done06}. Hence
we use the full Compton continuum shape as approximated by the {\sc
nthcomp} \citep{zdzi96, zyck99} model in {\sc xspec} to relate the energy flux 
in the spectrum to photon number for a given seed photon temperature.

The photon rate required to make the Comptonized spectrum depends also
on the optical depth $\tau$ of the corona, as only a factor
$(1-e^{-\tau})$ of the photons are scattered into the tail. This
depends on geometry as well as spectral index and electron
temperature, and requires a fully relativistic treatment (rather than
analytic approximations) for electron temperatures which are a
substantial fraction of the electron rest mass as assumed here. Hence
we use the {\sc xspec} model {\tt compps} \citep{pout96}
which includes all these factors. We choose a a spherical geometry
with isotropic seed photons (geom=-4), where spectral indices of
$\Gamma=2,\,2.5,$ and $3$ correspond to $\tau=0.85,\,0.4,$ and $0.2$ for
$kT_{\rm e}=100$~keV.  We use $T_{\rm BB}=0.2$~keV \citep{jin13}, but the
derived values for $\tau$ do not depend much on this parameter. 
Hence the required seed photon rate, $N_{\rm seed}$, needs to be larger than
the rate of photons in the Compton spectrum, 
$N_0$, by the factor $(1-e^{-\tau})^{-1}$, so $N_0=N_{\rm seed}(1-e^{-\tau})$.

These X-ray photons illuminate the disk, so can be either reflected or
reprocessed (thermalized) in the disk. The fraction which thermalize,
$\eta_{\rm th}$ can add to the seed photon flux from the intrinsic disk
emission. Photons from the lamp-post emitted into a solid angle
$\dd\Omega_{\rm L}$ illuminate a surface area perpendicular to the light rays
at the disk of $\dd S_{\dd\bot}$ at radius $R$ 
\citep[see geometry in][]{dovc14}.  This corresponds to a surface area on the 
disk $\dd S_\dd=2\pi
R^2 dR$ where $\dd S_{\dd\bot}/\dd S_\dd=H/D$ and $D^2=R^2+H^2$. Then the
illuminating flux on the disk is $F_{\rm ill}(R)=L_{\rm X}/(4\pi D^2)\times
\dd S_{\dd\bot}/\dd S_\dd =L_{\rm X}/(4\pi) H/D^3$. This leads to a rise in
temperature such that
$T_{\rm tot}^4(R)=F_{\rm ill}(R)/\sigma_{SB}+T^4(R)$. However, this increase
is negligible where the ratio $L_{\rm X}/L_{\rm disk}$ is small.

This is a somewhat surprising result in the context of the X-ray
spectrum, where the lamp-post model generates an extremely centrally
concentrated emissivity. However, even in a fully relativistic
treatment, this concentrated illumination is small compared to the
much more luminous disk emission.

\begin{figure}
    \epsfig{figure=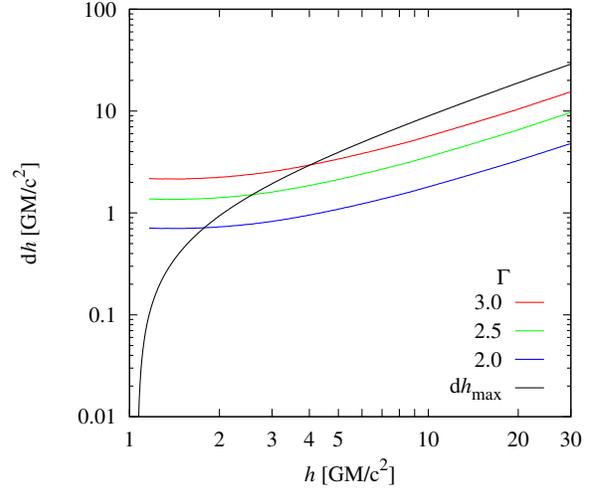,width=8cm}
    \caption{Newtonian radius of corona. 
      A size scale for the corona in Newtonian approximation required to
      produce $L_{\rm X}/L_{\rm disk}=0.02$ as observed at infinity for a 
      typical NLS1 ($10^7$\msun, $L_{\rm disk}=L_{\rm Edd}$) for 
      $\Gamma=2,\,2.5,$ and
      $3$ (bottom to top; corresponding to $\tau=0.85,\,0.4,$ and $0.2$,
      respectively) assuming
      $a=0.998$. More seed photons are required for softer spectra as
      the source only scatters a fraction $1-e^{-\tau}$.  The photon
      rate intercepted by a source on axis increases for decreasing
      $h$ down to $h\sim r_{\rm max}=1.6$, so the required source size
      decreases with $h$ and then flattens for $h<r_{\rm max}$ where the
      intensity is constant for isotropic emission. The black line
      shows the maximum radius where the source fits above the
      horizon.}
\label{fig:newton}
\end{figure}

The source at the height $H$ on the disk axis with a cross-section 
$\dd S_{\rm L} = \pi(\dd H)^2$ receives photons from the disk area $\dd S_\dd$ 
into the solid angle 
$\dd\Omega_{\rm L} = (\dd\Omega_{\rm L} / \dd S_\dd)\,\dd S_\dd = 
H/D^3\,2\pi R\,\dd R$. Thus the fraction of photons
from the disk which illuminate the source, 
$\dd S_{\rm L} \times \dd\Omega_{\rm L}$, 
increases with decreasing $R$ down to $R=H/\sqrt{2}$ where it reaches a maximum 
and then decreases with the further decrease of $R$.  
The rate at which photons are emitted from the disk at radius $R$ depends on the
temperature, 
$\dd N_{\rm disk}(R) \sim T^3(R)$, and the total rate of the intercepted photons 
by the source is
$\dd N_{\rm seed}=\dd N_{\rm disk}(R) \times \dd S_{\rm L} \times 
\dd\Omega_{\rm L}$ 
assuming that the source emits isotropically.
The total seed photon rate
$N_{\rm seed}=\int_R \dd N_{\rm seed}$, giving the radius of the corona, 
$\dd H$, from the constraint that $N_{\rm seed}$ needs to be such that it can 
make the Comptonized emission.

The seed photons have a range in temperature as they are produced over
a range of radii. However, this multi temperature seed photon
distribution has a clear peak at energy $E_0$, and can be fairly well
approximated by a single blackbody at a temperature $T_{\rm BB}=E_0/2.82$.
This seed photon temperature is always somewhat smaller than the peak
disk temperature, $T_{\rm max}$, by an amount that depends slightly on the 
source height. For $h\lesssim r_{\rm max}$ this reaches a maximum of $\sim
0.75T_{\rm max}$, while for $h\gg r_{\rm max}$ it asymptotes to 
$0.25T_{\rm max}$.

The black, green and blue lines in Figure~\ref{fig:newton} show the
derived source size as a function of height for $\Gamma=2,2.5$ and $3$
for a black hole mass of $10^7$\msun, $a=0.998$, $f_{\rm col}=2.4$ with
$L_{\rm disk}=L_{\rm Edd}$.  The seed photon density on axis increases with
decreasing $h$ until $h\sim r_{\rm max}\sim 1.6$ for our assumed
$a=0.998$, so the source size decreases up to this point, then tends
to a constant value of a few $R_{\rm g}$ as the flux no longer rises due to
the inner hole to the disk. We repeated the calculation for
$10^6$\msun, and for a color temperature correction of both $1$ as
well as $2.4$, but these parameters make a negligible difference to
the results. Our results are also fairly insensitive to our assumption
that $L_{\rm disk}=L_{\rm Edd}$, as it is $L_{\rm X}/L_{\rm disk}$ which is
important. With this set to $0.031$ then $\Gamma$ is the main
determinant of source size. Steeper spectra require more photons to
make the same luminosity, and also intercept a smaller fraction of the
available seed photons due to their lower optical depth. Both factors
increase the seed photons required by around a factor three $\Gamma=3$
compared to $\Gamma=2$, so the source size is a factor $\sqrt{3\times
3}=3\times$ larger for $\Gamma=3$ than $\Gamma=2$. Thus the source
size changes from $\sim 1-3R_{\rm g}$ as $\Gamma$ increases from $2-3$.  The
source is compact, as seems reasonable in view of the fast variability
and micro-lensing constraints, but not tiny, as also seems reasonable
in view of the energy emitted from the region.

\section{Fully relativistic approach}

\begin{figure}
    \epsfig{figure=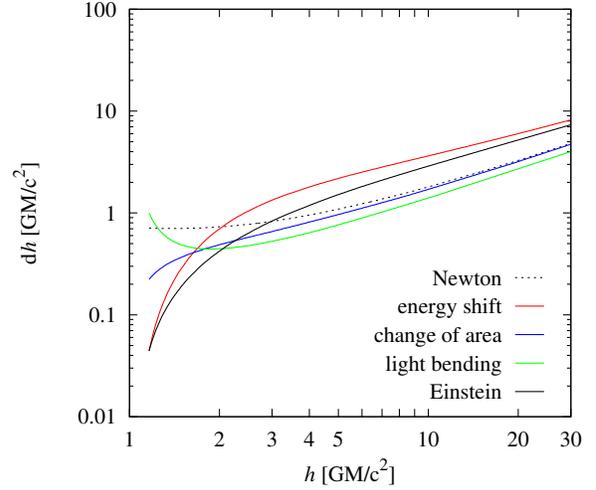,width=8cm}
    \caption{Contributing relativistic effects.
      A comparison of the size of each relativistic effect with
      the Newtonian results in Figure~\ref{fig:newton} for $\Gamma=2$
      (black dotted line). The major effect is from the energy shift,
      $g$, between the comoving disk and lamp-post frame as the photon
      rate is proportional to $g^3$ (red line). The change between
      coordinate distance and proper distance is also a factor (blue
      line), and these two effects are larger than light bending with
      aberration (green line). The black solid line shows the results
      including all the relativistic effects.}
\label{fig:releff}
\end{figure}

There are clearly a number of relativistic factors which affect this
calculation. We illustrate the effect of each of these in turn in
Figure~\ref{fig:releff}, where the dotted black line shows the
$\Gamma=2$ results from Figure~\ref{fig:newton}. Firstly the energy of
the seed photons from each annulus in the disk as seen in the lamp-post
frame is $kgT(r)$ where $g=E_{\rm L}/E_\dd$ is the energy shift between the
comoving disk frame and the lamp-post.  For $h>r_{\rm max}$ then this is a
redshift, while for $h<r_{\rm max}$ it is a blueshift. The photon rate
received at the lamp-post depends on $g^3$, so the required source size
increases for $h>r_{\rm max}$, and decreases for $h<r_{\rm max}$ (red line:
Figure~\ref{fig:releff}). Note, that $r_{\rm max}$ in relativistic case is 
shifted to higher values, since the maximum temperature as perceived by the 
corona is ${\rm max}[g(r)\,T(r)]$ instead of just $T_{\rm max}={\rm max}[T(r)]$.

Another factor is that the curvature of space-time means that a
source of given proper size extends over a smaller radial coordinate
distance as the source approaches the black hole. The radial
coordinate distance is important as this sets the position of the
horizon, $r_{\rm H}$, so that $\dd h< h-r_{\rm H}$ (Boyer Lindquist 
coordinates).
This effect means that the source radius in Boyer-Lindquist 
coordinates is always smaller than from the non-relativistic approach
(blue line: Figure~\ref{fig:releff}). 

The green line in Figure~\ref{fig:releff} shows the effect of
light bending and aberration on the solid angles. We use the code of
\citealt{dovc14} to follow the full photon geodesics, and
transform the angles discussed above. However, this is not such an
important effect. The solid black line in  Figure~\ref{fig:releff}
including all  relativistic effects is dominated by the 
boosting/de-boosting of the thermal emission. 

\begin{figure}
    \epsfig{figure=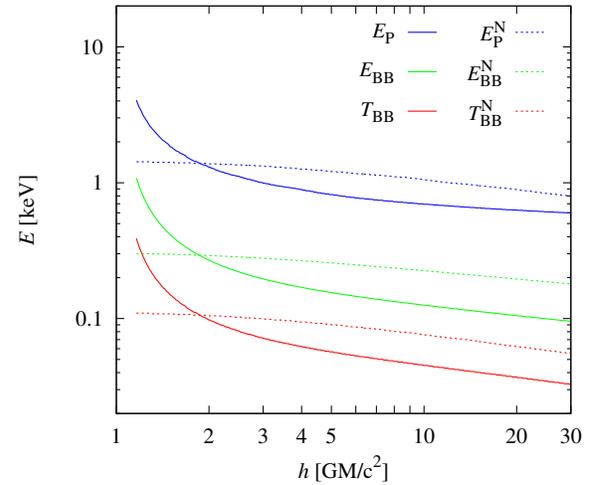,width=8cm}
    \caption{Photon temperature and average energy.
      The seed photon temperature, $T_{\rm BB}$ (red), average seed
      photon energy, $E_{\rm BB}$ (green), and average Comptonized photon
      energy, $E_{\rm p}$ (blue) evaluated in the lamp-post frame at different
      heights. The 'Newtonian' values are also shown (dotted lines).
}
\label{fig:e0}
\end{figure}

Figure~\ref{fig:e0} shows resulting fully relativistic mean energy of
seed photons and Comptonized photons in the lamp-post frame.  The red
solid line shows the seed photon temperature $T_{\rm BB}=E_0/2.82$, where
$E_0$ is the peak energy of the multicolor black
body received at the lamp-post. The green solid line shows the mean
seed photon energy, 
$E_{\rm BB}= \int EN_{\rm seed}(E) dE/\int N_{\rm seed}(E) dE$
from this multicolor blackbody. The fact that it is close to 
$2.82T_{\rm BB}$ shows that a single blackbody is a good approximation
to the shape of the seed photon distribution. The blue solid line
shows the mean energy of photons in the Comptonized spectrum for the
fiducial model with $\Gamma=2$ shown in Figure~\ref{fig:releff}.  The
dotted lines in Figure~\ref{fig:e0} show the corresponding Newtonian
energies. The change in energy is much larger in the fully
relativistic treatment due to the boosting/de-boosting of photons
between the co-moving disk frame and the lamp-post. Thus its impact on
the number of photons required in the Compton spectrum is larger than
in the 'Newtonian' case. However, the power extracted from the
electrons $L_{\rm e}=L_{\rm X}-N_0E_{\rm BB} = (E_{\rm p}-E_{\rm BB})N_0$ so it 
does not depend much on height as $E_{\rm p}$ is set by $E_{\rm BB}$, and both
change together for a given spectral index (see Figure~\ref{fig:e0}), so
this is canceled out by the opposite change in $N_0$. Thus our model is 
consistent with the constant dissipation in the intrinsic 
X-ray source which is assumed in the lamp-post geometry.
It comes out that about 
$80\%, 55\%$, and $40\%$ of the intrinsic primary X-ray luminosity needs to be
extracted from electrons for $\Gamma=2.0,\, 2.5$, and $3.0$, respectively. 

\begin{figure}
    \epsfig{figure=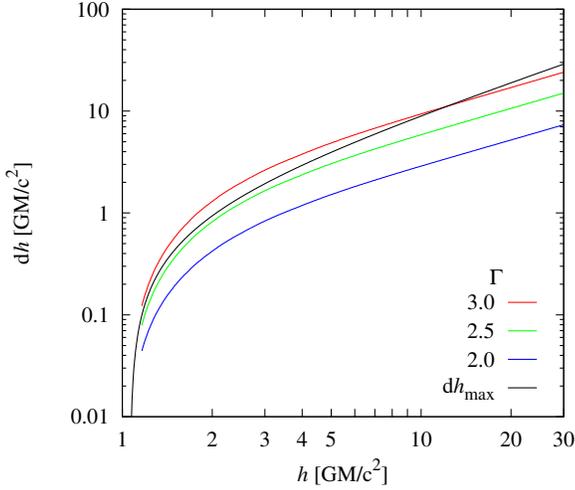,width=8cm}
    \caption{Corona radius in full general relativity.
      As for Figure~\ref{fig:newton}, but accounting properly for all general 
      relativistic effects. The radius
      of the source is in Boyer-Lindquist coordinates so the maximum
      radius for a source at height $h$ which can fit above the
      horizon at $r_{\rm H}$ is $h-r_{\rm H}$, (black line). 
A steep spectrum ($\Gamma=3.0$: red) on the spin axis does not
intercept enough flux from the disk to produce the observed
Comptonized emission for $h<12$.}
\label{fig:radius}
\end{figure}

Figure~\ref{fig:radius} shows the resulting minimum radial size estimate
of the corona in full general relativity, again for $\Gamma=2,\,2.5,$ and
$3$.  As before, softer spectra ($\Gamma=2$: blue; $2.5$: green; $3$
red) require a larger source size.  The black line shows the maximum
size, set by the constraint that the corona does not extend below
the horizon.  A source with $\Gamma=2$ and $2.5$ can have $\dd h\lesssim
h-r_{\rm H}$ at any height for an intrinsic luminosity of
$L_{\rm X}=0.031L_{\rm disk}$. However, there are not enough photons on the spin
axis of the black hole to make the softest spectra observed in NLS1 for a
source of this luminosity at a height $<12R_{\rm g}$.

\section{Application to 1H0707-495}

\begin{figure}
    \epsfig{figure=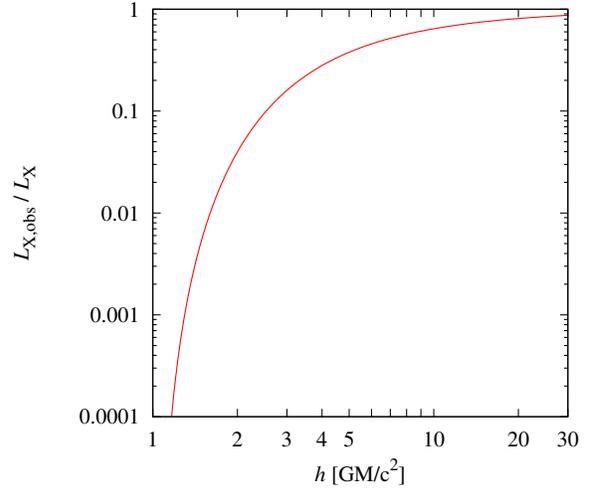,width=8cm}
    \caption{Flux decrease.
      The suppression of the intrinsic power law flux received at infinity as 
      a function of height.}
\label{fig:flux}
\end{figure}

\begin{figure}
    \epsfig{figure=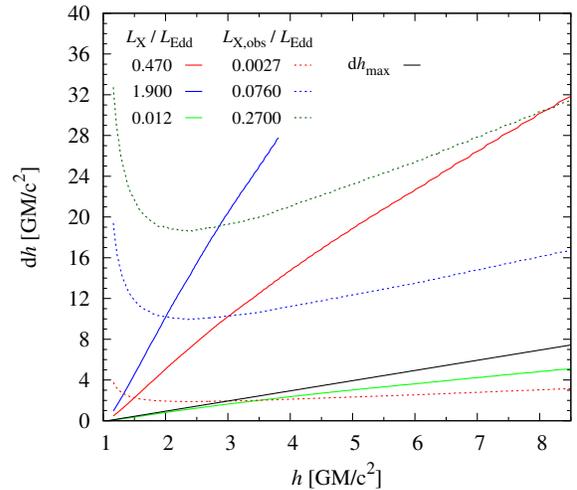,width=8cm}
    \caption{Size of the corona for 1H0707-495. {\em Solid curves} show the 
    required radius of the spherical patch of corona for a given intrinsic 
    luminosity. In pure light-bending scenario with constant intrinsic 
    luminosity the patch of corona would have to change its size along these
    graphs as it changes its position on the axis. {\em Dotted curves} show the
    corona radius for a given observed luminosity if the primary source is 
    located at certain height. Corona receives less thermal disk photons if it 
    is farther away from the disk thus the radius increases with larger heights. 
    However, for very small heights large intrinsic luminosities are needed
    for the same observed luminosity (see Figure~\ref{fig:flux}) and thus the
    required radius increases rapidly.}
\label{fig:1h0707}
\end{figure}

1H0707-495 is one of the most extreme NLS1 in terms of the inferred
relativistic effects \citep[][hereafter Z10]{fabi09,zogh10}. 
The 0.3-10~keV power law flux in these data from 2008 changes
from $\sim 2\times 10^{-13}$ to $2\times
10^{-11}$~erg~cm$^2$~s$^{-1}$ (Z10), and the lowest flux episode is similar
to the spectrum seen in 2011 where the inferred line emissivity has
index of $8.8$, requiring a source at $h<1.5$ (\citealt{fabi12},
hereafter F12; see also \citealt{daus12,dovc14}). 

Hence we use $h=1.5$ as the baseline for the lowest flux observation
in 2008.  The power law flux at each frequency is suppressed by the
factor $g_{\rm L}^2 \dd\Omega_{\rm L}/\dd\Omega_{\rm obs}$ where 
$g_{\rm L}=E_{\rm obs}/E_{\rm L} = \sqrt{
1-2h/(h^2+a^2)}$, and $\dd\Omega_{\rm L}/\dd\Omega_{\rm obs}$ is the 
light-bending term.  At such a low height, $g_{\rm L}=0.275$ (corresponding to 
redshift $z=2.632$), while $\dd\Omega_{\rm L}/\dd\Omega_{\rm obs}= 0.076$. 
Figure~\ref{fig:e0} shows that the seed photon temperature 
$T_{\rm BB}\sim 0.15$~keV in the
lamp-post frame, so we redshift a Compton spectrum with $\Gamma=3$ and
this seed photon temperature to find the bolometric correction from
the observed 0.3-10~keV bandpass is a factor of 2.32, so
$F_{\rm X,obs}= 4.64\times 10^{-13}$~erg~cm$^2$~s$^{-1}$, or
$L_{\rm X,obs}=0.0027L_{\rm Edd}$. Since this is suppressed by the factor
$g_{\rm L}^2 \dd\Omega_{\rm L}/\dd\Omega_{\rm obs}\sim 0.0057$, the intrinsic 
source flux in the lamp-post frame is $F_{\rm X}=8.14\times
10^{-11}$~erg~cm$^2$~s$^{-1}$, corresponding to $L_{\rm X}=0.47L_{\rm Edd}$.
This is substantially larger than the flux assumed in Section 3, so
all radii in Figure~\ref{fig:radius} will be increased by a factor $\sim
\sqrt{0.47/0.031}\sim 4$. These source sizes cannot fit within $h=10$,
let alone $h=1.5$ for any $\Gamma=2-3$.

We show the resulting source radius on a linear scale as the red solid
line in Figure~\ref{fig:1h0707}. It is always larger than the maximum
source size which can fit above the horizon (black solid line).  By
contrast, the red dotted line shows the source size which gives the
observed $L_{\rm X,obs}=0.0027$. The minimum source height which is
consistent with the source fitting above the horizon is
$h=3$. However, at this height, relativistic effects are not so
extreme and the emissivity radial power-law index is $\lesssim 4$ 
\citep{daus13}. 

The maximum X-ray flux from the source is a factor $100\times $
larger, implying $L_{\rm X,obs}=0.27L_{\rm Edd}$ (including the bolometric
correction). This flux change can be produced by moving the source up
to $h=8.2$ (see Figure~\ref{fig:flux}), but the dark green dotted line in
Figure~\ref{fig:1h0707} shows that a source of this observed luminosity
cannot fit anywhere on the spin axis for $h<9$.

However, the lowest luminosity power law has large uncertainties
(e.g. Z10, F12). Instead, we recalculate
our results from the mean spectrum from 2008, where the power law
detection is much more robust with a 0.3-10~keV flux of $6.2\times
10^{-12}$~erg~cm$^2$~s$^{-1}$ (F12). This has an emissivity index of
6.6, which still requires a very small lamp-post height of $h<1.5-2$
\citep{daus13,dovc14}.  We use the largest source
height of $h=2$, but this is probably an overestimate as seen by the
large reflected fraction of the data. Z10 quote a ratio of reflected
to power law flux of 1.3 in their 0.3-10~keV bandpass, but this
requires a solid angle of $\sim 10-20$ for their low ionization
parameter. The predicted solid angle for a lamp-post source
at $h=2$ is $\sim 7$ \citep{daus14}.

Thus the observed reflection dominance and centrally concentrated
emissivity are consistent with the lamppost at $h\sim 2$. 
This gives $g_{\rm L}=0.446$ ($z=1.240$), and a bolometric 
correction of $2.15$, hence a flux as seen at infinity of
$F_{\rm X,obs}=1.33\times 10^{-11}$~erg~cm$^2$~s$^{-1}$, corresponding to
$L_{\rm X,obs}=0.076L_{\rm Edd}$. This is suppressed by a factor $0.04$
(Figure~\ref{fig:flux}), so the intrinsic luminosity in the lamp-post
frame is $L_{\rm X}=1.9L_{\rm Edd}$. This implies an even larger source size
than before, with $\dd h\sim 10$ as shown by the blue solid line in
Figure~\ref{fig:1h0707}. 
Even the observed $L_{\rm X,obs}=0.076$ is too large
a luminosity to be produced by a source on the spin axis for any $h\le
9$ (blue dotted line).

The large intrinsic X-ray luminosities required in order to produce
the observed reflected emission dramatically exceed those
observed due to the strong suppression of the X-ray flux from the
combined effects of redshift and light bending.  Such large intrinsic
luminosities require a correspondingly large source size in order to
intercept sufficient seed photons. This size is inconsistent with the
size scale required for the source to fit above the horizon.

The largest intrinsic luminosity which is consistent with
the minimum source size which can fit above the horizon at small $h$
is shown by the solid green line in Figure~\ref{fig:1h0707}, which has
$L_{\rm X}=0.012L_{\rm Edd}$. This is a factor of 10 smaller than the {\em
observed} 0.3-10~keV flux at maximum, so this cannot possibly be a
viable solution. The corona would have to change its size with height by much
larger amount to increase the intrinsic primary flux to reach these observed
values and then it will no longer fit above the horizon (the green 
solid line in Figure~\ref{fig:1h0707} would have to curve up to join the 
dark-green dotted line).

\section{Discussion and Conclusions}

There is no viable 'lamp-post' source geometry (small source on the
black hole spin axis) which can explain the observations of
1H0707-495. The mismatch is most marked by considering the mean
spectrum from 2008 (Z10; F12), where the intrinsic power law is
clearly seen along with highly relativistically smeared reflection.  A
source which is at small enough height to explain the observed highly
centrally peaked emissivity and reflection dominance has $h\lesssim
2$.  However, the power law suppression from both light bending and
redshift here is extreme, and requires that the observed rather
small X-ray luminosity must instead represent a much larger intrinsic
X-ray luminosity, of $1.9L_{\rm Edd}$!  
This requires a minimum source
size of $\sim 10$ assuming that the power law is formed by Compton
up-scattering of seed photons from the disk, so it cannot have an
effective mean height of $h=2$, so it cannot reproduce the implied large 
reflected fraction, nor the extremely centrally peaked emissivity. 

A small source at  $h=2$ could be recovered if there is an additional source of
seed photons, perhaps from a strong magnetic field. A magnetic field is
surely required in order to 
produce the lamppost source in the first place, so it is worth exploring this 
possibility. To have the magnetic energy density be of the order of the X-ray 
energy density for $L_x\sim 2L_{Edd}$ requires
$B\sim 10^6$~G, but self-absorption will strongly limit the number of 
cyclo-synchrotron seed photons which can be produced \citep{dima97}. 

Nonetheless, the source could easily power both the large X-ray luminosity
and sustain a large magnetic field as the 
optical/UV emission from the outer disk in 1H0707-495 implies that the 
source is accreting at $\sim 200L_{Edd}$ for a maximally rotating black hole
with mass of $2\times 10^6M_\odot$ \citep{done15}. Thus it is energetically
possible that the very weak observed X-rays are due to a very high 
intrinsic X-ray flux which is  dissipated so close to the black hole that it
is strongly suppressed.  However, $L_{\rm X}>L_{\rm Edd}$ is a completely new
requirement for NLS1. Generically these are observed to be X-ray weak
even in sources where the spectra are not dominated by reflection, and
the observed reflection is not strongly smeared 
\citep[e.g.][]{jin12}.

A tiny lamp-post corona geometry cannot explain all the aspects of the
X-ray data: the source must be somewhat extended. This highlights an
issue with our calculations as our model simply scales the seed photon
density on the spin axis to set the source size, without explicitly
re-calculating the seed photon density for the off-axis extent of the
source and for the expected rotation of such a source. 
This is clearly required, and means that the source will
intercept more seed photons from the disk as the black hole spin axis
minimizes the photon density. Thus a full 3D computation is required to
estimate the actual size and shape of the corona. Calculations of 
extended source geometries have been done by  e.g. 
\citealt{wilk12,wilk14,daus13,daus14}. However, it seems highly unlikely that 
the reflecting disk can remain thin and flat for such a super Eddington flow, so
the reflector geometry as well as the coronal size and shape should also be 
considered. 

\bigskip
\section*{Acknowledgments}
The research leading to these results has received funding from the European 
Union Seventh Framework Programme (FP7/2007-2013) under grant agreement 
n$^\circ$312789. CD acknowledges STFC grant ST/L00075X/1
and thanks the FP7 strong gravity consortium for funding a visit 
to Prague where much of the work was done. 

\appendix

\section{Multi-color black body received by the corona}

The disk emits with standard Novikov-Thorne emissivity, so
forms a blackbody with temperature $T(r)$ from each radius $r$. 
We allow for the possibility of a color temperature correction, which
increases the temperature by a factor $f_{\rm col}$, and decreases the
normalization by $f_{\rm col}^4$ so as to give the same total luminosity.

Photons from the disk are emitted from a Boyer-Lindquist area 
$\dd S_\dd=2\pi r\dd r$ at energy $E_\dd$ in the local frame
comoving with the disk. These are received at the lamp-post with energy
$E_{\rm L}$ (such that $g=E_{\rm L}/E_\dd$) from a solid angle $\dd\Omega_{\rm
L}$. The multi-color black body radiation received at the lamp-post from the
accretion disk is $F_{\rm in}\equiv \dd N/(\dd t\,\dd S_{\rm L})$, i.e. 
the photon number density flux received at the lamp-post with a perpendicular 
cross-section $\dd S_{\rm L}$. Both $\dd S_{\rm L}$ and $\dd\Omega_{\rm L}$ are 
measured locally in the reference frame of the lamp-post. The incoming thermal
photon flux has to be integrated over the disk,

\begin{eqnarray}
\nonumber   
F_{\rm in} & = & 2\pi\int_{0}^{\infty}\dd E\int_{r_{\rm in}}^{r_{\rm out}} 
\dd r\,r\,\frac{2}{f_{\rm col}^4\,h^3\,c^2} \frac{E^2}{e^{\,E/kgT}-1}\,
\frac{\dd\Omega_{\rm L}}{\dd S_\dd}
\\[2mm]
& = & 2\pi\frac{4\zeta(3)k^3}{f_{\rm col}^4\,h^3\,c^2}
\int_{r_{\rm in}}^{r_{\rm out}} 
\dd r\,r\,\frac{\dd \Omega_{\rm L}}{\dd S_\dd}\,(gT)^3\, ,
\\[2mm]
T & = & T_{\rm norm}\,r^{-3/4} 
\left[\frac{\mathcal{L}(r)}{\mathcal{C}(r)}\right]^{1/4}
\left(\frac{\dot{M}}{M_\odot\,y^{-1}}\right)^{1/4} 
\left(\frac{M}{M_\odot}\right)^{-1/2}\, ,
\\[2mm]
T_{\rm norm} & \equiv & f_{\rm col}\,c^{3/2}
        \left(\frac{3}{8\pi G^2 M_\odot y\,\sigma}\right)^{1/4}\, ,\\[2mm]
\nonumber
\end{eqnarray}
where all symbols have their usual meanings,
$\zeta(3)\approx 1.202056903159594$ is an Ap\'{e}ry's constant,
and where $T(r),
{\mathcal{L}(r)},{\mathcal{C}(r)}$ are given in e.g. \cite{novi73} or 
\cite{page74}.

\section{Disk temperature increase due to illumination}

The accretion disk is illuminated with the energy flux
\begin{eqnarray}
F_{\rm inc}(r) & = & g_{\rm L}\,g\,\frac{\dd\Omega_{\rm L}}{\dd S_\dd}
\frac{L_{\rm X}}{4\pi}
\end{eqnarray}
where we used the local area on the disc to be 
$\dd S = \dd S_{\rm d}/(g_{\rm L}\,g)$
\citep[see e.g. eq.~(2.8) in][]{dovc04} with $g_{\rm L}=E_{\rm obs}/E_{\rm L}$
and $g_{\rm L}\,g=E_{\rm obs}/E_{\rm d}$.
Thus temperature will rise due to the illumination to
\begin{equation}
T=\left(T_{\rm BB}^4+\frac{\eta_{\rm th}\,F_{\rm inc}}{\sigma}\right)^{1/4}
\end{equation}
with the $\eta_{\rm th}$ being the fraction of the illumination flux
that is thermalized. For NLS1, the ratio of power in X-rays versus the
intrinsic disk emission is small, so re-processing makes only a negligible
effect on the photon flux. 

\section{Note on the derivation of cross section in Boyer-Lindquist coordinates}

We define the cross section in the Boyer-Lindquist coordinates as
\begin{equation}
\label{eq:BL-area}
\dd S_{\rm BL} = n^\beta \dd^2S_{t\beta}\,,
\end{equation}
where the unit vector $n^\beta=(0,p^i)/\sqrt{p^ip_i}=-(0,p^i)/
(p_\mu U_{\rm L}^\mu)$ has the direction of the
3-vector of the photon momentum. We will use the fact that the cross section
is the same for all observers \citep[see e.g.  eq.~(2.6) in][]{dovc04}\footnote{
Here we would like to emphasize that while this equation holds true for any 
4-velocity, $U^\alpha$, i.e. for all time-like vectors, it is not valid for all
space-like vectors. Indeed, if we use e.g. a vector $X^\alpha$ that lies in 
the area defined by $\dd^2S_{\alpha\beta}$ so that the first term in the equation
is zero, the second term will be non-zero if the scalar product 
$p_\alpha X^\alpha$ is non-zero. Thus the 4-vector in the parenthesis of the 
eq.~(\ref{eq:cross-section}) is not zero as is claimed in the eq.~(2.7)
in \cite{dovc04}.}
\begin{equation}
\label{eq:cross-section}
(p^\beta\,\dd^2S_{\beta\alpha}-p_\alpha\,\dd S^\perp)\,U^\alpha = 0
\end{equation}
for the time-like 4-vector $U^\alpha=(1,0,0,0)$. In our case the locally 
observed area is perpendicular to the photon 3-momentum, 
$\dd S_{\rm L}=\dd S^\perp$, and further $p_t=-1$. Thus we get
\begin{equation}
\label{eq:area_ratio}
\dd S_{\rm BL} = -\frac{p^\beta \dd^2S_{t\beta}}{p_\mu U_{\rm L}^\mu}=
g_{\rm L}\dd S_{\rm L}\,.
\end{equation}
In our calculations we use the radius, $\dd h$, of the Boyer-Lindquist proper 
cross-section from assuming the circular shape,
$\dd S_{\rm BL} = \pi \dd h^2$. The radius defined in this way does not equal to 
the change in Boyer-Lindquist coordinates. In fact the proper area that would be 
parallel or perpendicular with the axis would depend on the change of the BL
coordinates, $\dd h$, as $\pi\dd h^2 \sqrt{h^2+a^2}/h$ or 
$\pi\dd h^2 \sqrt{(h^2+a^2)(h^2-2h+a^2)}/h^2$, respectively. This would lead to
$9\%$ smaller radii in the first case and $59\%$ larger radii in the second
case for the spin $a=0.998$ and height $h=1.5$. Thus our estimation of corona
radius from proper area will not differ too much from the radius that would be
given in the Boyer-Lindquist coordinates.

\section{The intrinsic X-ray flux emitted by the corona}

We assume the local X-ray emission of the corona to be given by the 
non-thermal Comptonization model, {\sc nthcomp} \citep{zdzi96, zyck99}, 
available in the spectral fitting package {\sc XSPEC} \citep{arna96}. 
We compute the {\sc nthcomp} normalization, $N_{\rm L}$, from the observed 
photon flux
\begin{eqnarray}
\nonumber
F_{\rm X,obs}(0.3-10{\rm keV}) & = & \frac{1}{4\pi D^2}L_{\rm X,obs}(0.3-10{\rm keV})=
\frac{g^2_{\rm L}}{4\pi D^2} \frac{{\rm d}\Omega_{\rm L}}
{{\rm d}\Omega_{\rm obs}} L_{\rm X}(0.3/g_{\rm L}-10/g_{\rm L}{\rm keV})=\\
& = & \frac{g^2_{\rm L}}{4\pi D^2}\frac{{\rm d}\Omega_{\rm L}}{{\rm d}\Omega_{\rm obs}} 
N_{\rm L}\int_{0.3/g_{\rm L}}^{10/g_{\rm L}}E\;{\tt nthcomp}(E)\;{\rm d}E\,.
\end{eqnarray}
Here we needed to use the computation of {\sc nthcomp} luminosity in the corona 
frame since it depends on the temperature of the incident black-body radiation.
Then the total observed luminosity, $L_{\rm X,obs}$, and the total intrinsic 
photon production rate in the corona, $F_{\rm out}$, are
\begin{equation}
L_{\rm X,obs} = 4\pi D^2\,\frac{\int_{0}^{\infty}E\;{\tt nthcomp}(E)\;{\rm d}E}
{\int_{0.3/g_{\rm L}}^{10/g_{\rm L}}E\;{\tt nthcomp}(E)\;{\rm d}E}\,
F_{\rm X,obs}(0.3-10{\rm keV})
\end{equation}
and
\begin{equation}
F_{\rm out} = 4\pi D^2\,\left(g^2_{\rm L} \frac{{\rm d}\Omega_{\rm L}}
{{\rm d}\Omega_{\rm obs}}\right)^{-1}
\frac{\int_{0}^{\infty}{\tt nthcomp}(E)\;{\rm d}E}
{\int_{0.3/g_{\rm L}}^{10/g_{\rm L}}E\;{\tt nthcomp}(E)\;{\rm d}E}\,
F_{\rm X,obs}(0.3-10{\rm keV})\,,
\end{equation}
respectively.

\medskip
To complete basic equations leading to our corona size estimates we conclude
with the number of scattered photons that make up the intrinsic photon 
production rate,
\begin{equation}
(1-e^{-\tau})\,F_{\rm in}\,\dd S_{\rm L} = F_{\rm out}\,.
\end{equation}

\end{document}